%% file: preprint.tex
\documentstyle[aps,pra,twocolumn,floats]{revtex}

\begin{document}
\draft
\title{Determination of entangled quantum states of a trapped atom}

\author{S. Wallentowitz, R.L. de Matos Filho, and W. Vogel \\ 
  Arbeitsgruppe Quantenoptik, Fachbereich Physik, Universit\"at
  Rostock \\ Universit\"atsplatz 3, D-18051 Rostock, Germany}

\date{September 18, 1996} 

\maketitle

\begin{abstract}
  We propose a method for measuring entangled vibronic quantum states
  of a trapped atom. It is based on the nonlinear dynamics of the
  system that appears by resonantly driving a weak electronic
  transition.  The proposed technique allows the direct sampling of a
  Wigner-function matrix, displaying all knowable information on the
  quantum correlations of the motional and electronic degrees of
  freedom of the atom. It opens novel possibilities for testing
  fundamental predictions of the quantum theory concerning interaction
  phenomena.
\end{abstract}


\narrowtext 

\section{Introduction}
Entanglement is one of the most striking aspects of quantum
mechanics~\cite{Schroed,EPR}. In classical physics, two interacting
systems retain their individuality during the interaction process and
become completely independent of each other after their coupling has
been switched off. By way of contrast, quantum theory predicts a
completely different behaviour of interacting systems. When two quantum
systems are brought to interact, their identities become in the course
of time more and more entangled, so that a state-vector description of
each system is in general precluded. They build an entangled composite
system, whose state-vector cannot be separated into a product of the
states of the subsystems. More surprising, the entanglement is
preserved even after switching off the interaction, so that a
measurement on one system will affect the other one. The individual
systems can be characterised by (reduced) density matrices.  However,
even when the reduced density matrices of both subsystems are known,
important information on the physics of their interaction and the
related entanglement is lost.

Recently, entangled quantum states of an electronic system and a
harmonic oscillator have been realized with trapped
atoms~\cite{wineland} and in cavity QED~\cite{haroche-entangled}. The
signatures of the entanglement have been partially demonstrated by
quantum measurements. The determination of the full quantum
statistical information of such quantum states, however, requires new
types of quantum measurements. Moreover, it is interesting in this
context that with both types of systems the Jaynes--Cummings
model~\cite{jaynes-cummings} and its nonlinear trapped-atom
counterpart~\cite{vogel-matos} could be
realized~\cite{meekhof,haroche-jcm}. In the Jaynes--Cummings
interaction the quantum correlation between the interacting systems
has been predicted to show surprising features~\cite{knight}. In all
these cases methods for determining the full information on the
quantum state of entangled systems are desired for getting more
insight into interaction phenomena in the quantum world.

Several approaches have been proposed for determining the complete
information on the quantum state of single quantum systems. In
particular, homodyne tomography has been established for optical
fields~\cite{raymer}. Further simplifications of the method have been
introduced, including the direct statistical sampling of the density
matrix in field-strength (quadrature)
representation~\cite{kuehn-vogel-welsch,zucchetti,vowe-buch},
photon-number representation~\cite{dariano-leonhardt-co}, and
symplectic tomography~\cite{manko-family}. Tomographic methods have
also been realized for molecular vibrations~\cite{walmsley} and
proposed for reconstructing the motional quantum state of trapped
atoms~\cite{wallentowitz-vogel1}. Alternatively, phase-space
distributions can be determined by measuring the number statistics of
the quantum state of interest, after introducing appropriate coherent
displacements~\cite{wallentowitz-vogel2}. A method of the latter type
has recently been used to reconstruct the motional state of a trapped
atom \cite{leibfried}. For quantum systems undergoing a
Jaynes-Cummings type dynamics~\cite{jaynes-cummings}, such as high-$Q$
cavity fields~\cite{walther,haroche-jcm} or trapped
atoms~\cite{blockley,vogel-matos,meekhof}, various methods have been
considered to study the quantum statistics of the bosonic subsystem
(the field or the center-of-mass motion),
cf.~\cite{vogel,bardroff-schleich,bardroff-schleich2,leibfried}. All
these methods, however, give no insight into the entanglement of
quantum systems.

In this contribution we propose a method for determining entangled
vibronic quantum states of a trapped atom. In Sec.~\ref{sec:charac} of
this paper we introduce a Wigner-function matrix, which contains the
full information on the composite system under study.  After a brief
discussion of its properties we show that it can be determined via
measurements of entangled, motional number statistics. In the
following section we present a concrete scheme for measuring those
quantities. It relies on the nonlinearities appearing in the motional
dynamics of the trapped atom interacting with laser light, which have
been predicted for atomic localisations beyond the Lamb--Dicke
regime~\cite{vogel-matos} and confirmed in recent
experiments~\cite{meekhof}. By monitoring the electronic dynamics,
combined with coherent displacements of the motional subsystem, one
can perform a direct sampling of the Wigner-function matrix of the
vibronic state. In Sec.~\ref{sec:numerics} we present a numerical
simulation of the reconstruction of the Wigner-function matrix and
briefly discuss the practical aspects of our method. A summary and
some conclusions are given in Sec.~\ref{sec:summary}. 

\section{Characterisation of entangled quantum states}\label{sec:charac}
The motional quantum state of a trapped atom is usually described by a
density operator $\hat{\rho}$, which is an operator in the Hilbert
space ${\cal H}_{\rm cm}$ of the motion of the atom in the trap. To
get a phase-space description of the motional state of the atom one is
lead to quasiprobability distributions such as for example the Wigner
function. This function can be obtained from the density operator as
follows~\cite{glauber,vowe-buch}
\begin{equation}
  \label{eq:wigner1}
  W(\alpha) = \langle \hat{\delta}(\alpha\!-\!\hat{a}) \rangle = {\rm
    Tr} \left[ \hat{\rho} \, \hat{\delta}(\alpha\!-\!\hat{a}) \right]
  ,
\end{equation}
where the operator-valued delta-function
$\hat{\delta}(\alpha\!-\!\hat{a})$ is the Fourier-transform of the
displacement operator $\hat{D}(\alpha) \!=\! \exp (\alpha
\hat{a}^\dagger \!-\!  \alpha^\ast \hat{a})$ and reads
as~\cite{glauber}
\begin{eqnarray}
  \label{eq:Top}
  \hat{\delta}(\alpha\!-\!\hat{a}) & = & \frac{1}{\pi} \int \!
  d^2\!\xi \, \hat{D}(\xi) \, e^{\alpha \xi^\ast - \alpha^\ast \xi}
  \nonumber \\ & = & \frac{2}{\pi} \, \hat{D}(\alpha)
  (-1)^{\hat{a}^\dagger \hat{a}} \hat{D}^\dagger (\alpha) .
\end{eqnarray}
Equation~(\ref{eq:wigner1}) can be inverted to get the density
operator out of the Wigner function
\begin{equation}
  \label{eq:inversion1}
  \hat{\rho} = \int \! d^2\!\alpha \, W(\alpha) \,
  \hat{\delta}(\alpha\!-\!\hat{a}) ,
\end{equation}
so that the knowledge of $W(\alpha)$ is equivalent to the knowledge of
the motional density operator $\hat{\rho}$.

\subsection{Wigner-function matrix}
Even though the Wigner-function $W(\alpha)$ displays all the
obtainable knowledge about the motional state of the atom, it gives no
information on its electronic degrees of freedom. For a complete
description of the atomic state --- i.e. including the electronic
degrees of freedom --- one has to generalise the concepts leading to
the definition of the Wigner-function. For this purpose we introduce
the density operator $\hat{\varrho}$ of the whole system, which is now
an operator in the product-space of motional and electronic Hilbert
spaces ${\cal H}_{\rm cm} \otimes {\cal H}_{\rm el}$. The reduced
density operators for the motional and electronic subsystems can be
obtained by taking the trace over the Hilbert spaces of the electronic
system (el) and the center-of-mass motion (cm), respectively,
\begin{equation}
  \label{eq:subsytems}
  \hat{\rho} = {\rm Tr}_{\rm el} \left[ \hat{\varrho} \right] , \quad
  \hat{\sigma} = {\rm Tr}_{\rm cm} \left[ \hat{\varrho} \right] .
\end{equation}
We now define a Wigner-function matrix describing the complete quantum
state of the trapped atom by generalising Eq.~(\ref{eq:wigner1}),
\begin{equation}
  \label{eq:wigner2}
  W_{ij}(\alpha) = \langle \hat{\delta}_{ij}(\alpha\!-\!\hat{a})
  \rangle = {\rm Tr} \left[ \hat{\varrho} \,
    \hat{\delta}_{ij}(\alpha\!-\!\hat{a}) \right] ,
\end{equation}
where $\hat{\delta}_{ij}(\alpha\!-\!\hat{a})$ is now an operator in
the product-space ${\cal H}_{\rm cm} \otimes {\cal H}_{\rm el}$ and is
defined by
\begin{equation}
  \label{eq:Top2}
  \hat{\delta}_{ij}(\alpha\!-\!\hat{a}) = \hat{A}_{ji} \,
  \hat{\delta}(\alpha\!-\!\hat{a}) .
\end{equation}
Here we have used the electronic flip-operators $\hat{A}_{ji} \!=\!
|j\rangle \langle i|$, which describe transitions from the electronic
state $|i\rangle$ to the state $|j\rangle$. In close analogy with the
case of the motional sub-system~(\ref{eq:inversion1}), the density
operator of the composite system can be obtained from this
Wigner-function matrix as
\begin{equation}
  \label{eq:inversion2}
  \hat{\varrho} = \sum_{ij} \int \! d^2\!\alpha \, W_{ij}(\alpha) \,
  \hat{\delta}_{ji}(\alpha\!-\!\hat{a}) ,
\end{equation}
showing that the complete information on the quantum state of the atom
is contained in the Wigner-function matrix $W_{ij}(\alpha)$.

\subsection{Properties of the Wigner-function matrix}
Let us now discuss some properties of the Wigner-function matrix
$W_{ij}(\alpha)$. It is easy to see that its diagonal elements
$W_{ii}(\alpha)$ represent the Wigner functions of the conditioned
density operators $\hat{\varrho}_{ii}$ correlated to the electronic
state $|i\rangle$.  More precisely, this means that if one would
measure the electronic state of the trapped atom to be in state
$|i\rangle$, the corresponding (unnormalised) conditioned Wigner
function of the motional sub-system would be $W_{ii}(\alpha)$. The
norm of this conditioned Wigner function is simply the occupation
probability $\sigma_{ii}$ of the electronic state $|i \rangle$,
whereas the norm of the off-diagonal elements $W_{ij}(\alpha)$ ($i
\!\neq\! j$) represents the electronic coherence
\begin{equation}
  \label{eq:norm2}
  \int \! d^2\!\alpha \, W_{ij}(\alpha) = {\rm Tr} \left[ \hat{A}_{ji}
    \, \hat{\varrho} \right] = \sigma_{ij} .
\end{equation}
Moreover, from Eqs.~(\ref{eq:wigner2}) and (\ref{eq:Top2}) it is seen
that the Wigner-function matrix is hermitian
\begin{equation}
  \label{eq:hermitian}
  W_{ij}(\alpha) = W_{ji}^\ast (\alpha) ,
\end{equation}
so that in order to have the full information on the quantum state
under consideration, it is sufficient to know the real-valued diagonal
elements of the Wigner-function matrix $W_{ii}(\alpha)$ and real and
imaginary part of the off-diagonal elements $W_{ij}(\alpha)$
$(i\!<\! j)$.

If one measures the electronic state to be in an arbitrary
superposition
\begin{equation}
  \label{eq:superposition}
  |\psi\rangle = \sum_i \psi_i \, |i\rangle ,
\end{equation}
the conditioned density matrix of the motional sub-system is
\begin{equation}
  \label{eq:supdop}
  \hat{\rho}^{(|\psi\rangle)} = {\rm Tr}_{\rm el} \left[ |\psi\rangle
    \langle\psi | \hat{\varrho} \right] = \sum_{ij} \psi_i^\ast \psi_j
  \, \hat{\varrho}_{ij} .
\end{equation}
The Wigner function of this motional quantum state can now be
represented in terms of the Wigner-function matrix~(\ref{eq:wigner2})
\begin{equation}
  \label{eq:supwig}
  W^{(|\psi\rangle)}(\alpha) = \sum_{ij} \psi_i^\ast \psi_j \,
  W_{ij}(\alpha) .
\end{equation}
This result shows that the off-diagonal elements of the
Wigner-function matrix, which are in general complex-valued, contain
the information on the electronic coherences and the corresponding
motional states, which are entangled with these coherences. 

On the other hand, if one has no information about the electronic
state of the atom, the density operator of the motional sub-system is
given by
\begin{equation}
  \label{eq:redop}
  \hat{\rho} = {\rm Tr}_{\rm el} \left[ \hat{\varrho} \right] = \sum_i
  \hat{\varrho}_{ii} .
\end{equation}
The Wigner function of this reduced density operator reads as
\begin{equation}
  \label{eq:redwig}
  W(\alpha) = \sum_i W_{ii}(\alpha) ,
\end{equation}
and is simply the trace of the Wigner-function matrix.

In Figure~1 we show an example for the Wigner-function matrix of an
entangled state corresponding to the so called ``Schr\"odinger-cat''
states.
\begin{figure}
  \label{fig1}
  \begin{center}
    \vspace*{-1.5cm}\hspace*{-20mm}\input{wig.pstex_t}\vspace*{-15mm}
  \end{center}
  \caption{Wigner-function matrix $W_{ij}(\alpha)$ of the
    entangled vibronic quantum state $|\psi\rangle$ [cf.
    Eq.~(\ref{eq:ent})] with $\beta\!=\!2$.}
\end{figure}
The quantum state considered here has been recently prepared
experimentally with a trapped atom by Monroe et al.~\cite{wineland}.
It is a quantum superposition of two motional coherent states,
$|\pm\!\beta\rangle$, of amplitudes $\beta$ and $-\beta$, entangled
with the upper and lower electronic states, respectively. This state
can be given by the following expression
\begin{equation}
  \label{eq:ent}
  |\psi\rangle = \frac{1}{\sqrt{2}} \Big( \, |\beta\rangle |2\rangle -
  |-\!\beta\rangle |1\rangle \, \Big) .
\end{equation}
The Wigner-function matrix for this entangled quantum state can be
explicitely given by
\begin{eqnarray}
  \label{eq:wiganalytisch}
  W_{11}(\alpha) & = & \frac{2}{\pi} \exp(-2|\alpha+\beta|^2) , \\
  \nonumber W_{22}(\alpha) & = & \frac{2}{\pi}
  \exp(-2|\alpha-\beta|^2) , \\ \nonumber W_{12}(\alpha) & = &
  -\frac{1}{\pi} \, \exp(-2 |\alpha|^2) \exp[2i \, {\rm Im} (\alpha
  \beta^\ast )] .
\end{eqnarray}
It is clearly seen, that $W_{11}(\alpha)$ and $W_{22}(\alpha)$
represent the coherent states with amplitudes $-\beta$ and $\beta$,
respectively, whereas $W_{12}(\alpha)$ contains the information on the
electronic coherence and the quantum interference effects inherent in
the entangled state under study, see Fig.~1.

If the quantum state under consideration can be written as a product
state
\begin{equation}
  \label{eq:opproduct}
  \hat{\varrho} = \hat{\rho} \otimes \hat{\sigma} ,
\end{equation}
the corresponding Wigner-function matrix reads as
\begin{equation}
  \label{eq:wigproduct}
  W_{ij}(\alpha) = \sigma_{ij} W(\alpha) .
\end{equation}
States of this type contain no entanglement between the motional and
electronic degrees of freedom. The corresponding Wigner-function
matrix $W_{ij}(\alpha)$ is of the same shape for all indices $i,j$,
determined by the motional Wigner function $W(\alpha)$ and weighted by
the electronic density matrix elements $\sigma_{ij}$.

\subsection{Representation by displaced number statistics}
For determining the Wigner-function matrix $W_{ij}(\alpha)$ from
measured data we note that it can be represented in terms of matrix
elements of the coherently displaced density operator $\hat{\varrho}$,
where only the motional diagonal elements are needed. To show this we
make use of Eqs.~(\ref{eq:wigner2}), (\ref{eq:Top2}) together with
Eq.~(\ref{eq:Top}). By taking the trace over the electronic
sub-system, the Wigner-function matrix can be written as
\begin{equation}
  \label{eq:deriv1}
  W_{ij}(\alpha) = \frac{2}{\pi} \, {\rm Tr} \left[ \hat{\varrho}_{ij}
    \, \hat{D}(\alpha) \, (-1)^{\hat{a}^\dagger \hat{a}}
    \hat{D}^\dagger (\alpha) \right] ,
\end{equation}
where $\hat{\varrho}_{ij} \!=\! \langle i| \hat{\varrho} |j \rangle$
is still an operator acting on ${\cal H}_{\rm cm}$. By using the
cyclic property of the trace and performing the trace in terms of
number-states of the harmonic vibration in the trap, the
Wigner-function matrix reads as
\begin{equation}
  \label{eq:wigseries}
  W_{ij}(\alpha) = \frac{2}{\pi} \sum_{n=0}^\infty (-1)^n
  \varrho_{ij}^{nn}(\alpha) .
\end{equation}
Here we have made use of the coherently displaced density operator
\begin{equation}
  \label{eq:dispop}
  \hat{\varrho}(\alpha) = \hat{D}^\dagger(\alpha) \, \hat{\varrho} \,
  \hat{D}(\alpha) .
\end{equation}
For representing the Wigner-function matrix we only need the diagonal
elements of $\hat{\varrho}(\alpha)$ with respect to the motional
degree of freedom
\begin{equation}
  \label{eq:dispmatrix}
  \varrho_{ij}^{nn}(\alpha) = \langle n | \langle i | \,
  \hat{\varrho}(\alpha) \, | j \rangle | n \rangle .
\end{equation}
For convenience we will call these quantities displaced, entangled
number statistics.

To get the full information on the entangled quantum state, it is
sufficient to measure the displaced, entangled number statistics
(\ref{eq:dispmatrix}), in generalisation of the method presented in
Ref.~\cite{wallentowitz-vogel2,leibfried}. The initial coherent
displacement can be realized by applying a radio-frequency field, as
has been done in the experimental determination of the quantum state
of the motional subsystem~\cite{leibfried}. In the following we
present a measurement scheme for determining the displaced, entangled
number statistics $\varrho_{ij}^{nn}(\alpha)$.
 
\section{Measurement scheme}
The basic scheme consists in a three-level electronic system of
$V$-type which is superimposed by the energy levels of the motion of
the atom in the harmonic trap potential, see Fig.~2.
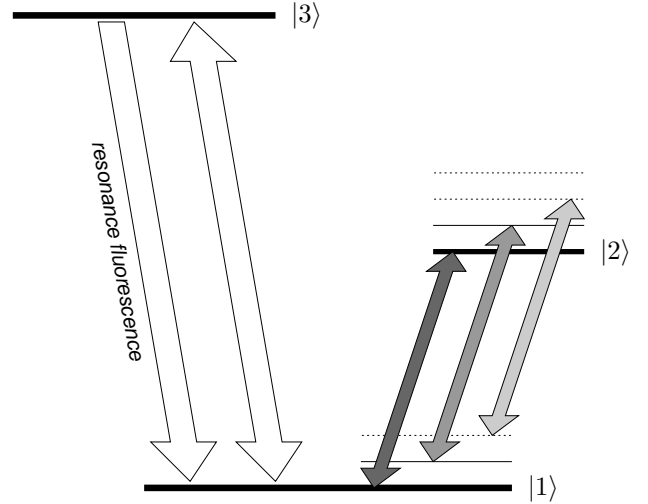
\begin{figure}
  \label{fig2}
  \begin{center}
    \vspace*{-2mm}
    \input{scheme.pstex_t}
    \vspace*{3mm}
  \end{center}
  \caption{Three-level electronic system of the trapped atom for the
    measurement of the displaced, entangled number statistics. The
    weak electronic transition $|1\rangle \leftrightarrow |2\rangle$
    is tested by probing the strong transition $|1\rangle
    \leftrightarrow |3\rangle$ for resonance fluorescence.}
\end{figure}
A weak electronic transition $|1\rangle \leftrightarrow |2\rangle$ is
the one of interest with respect to its entanglement with the motional
degree of freedom. The dynamics of this transition is monitored with
very high quantum efficiency by testing a strong, auxiliary transition
$|1\rangle \leftrightarrow |3\rangle$ for the appearance of resonance
fluorescence \cite{quantum-jump}. We assume that the weak transition
is driven in the resolved sideband limit. When the laser is tuned on
resonance with the vibrationless line, the corresponding interaction
Hamiltonian (in the interaction picture) reads
as~\cite{vogel-matos,matos-vogel-qnd}
\begin{equation}
  \label{hamiltonian}
  \hat{H}_{\rm int} = \frac{1}{2} \hbar \Omega \hat{f}
  (\hat{a}^\dagger \hat{a} ) \hat{A}_{12} + {\rm h.c.}
\label{Eq1}\end{equation}
Here $\Omega \!=\! |\Omega| \exp(i\varphi)$ is the Rabi frequency of
the laser interacting with the weak electronic transition and
$\varphi$ is the phase of the laser field~\cite{footnote1}. The
operators $\hat{a}$ and $\hat{a}^\dagger$ are the annihilation and
creation operators of a vibrational quantum, respectively. The
function $\hat{f} ( \hat{a}^\dagger \hat{a} )$, describing the
nonlinearities in the vibronic coupling, is given in its normally
ordered form by
\begin{equation}
  \hat{f} (\hat{a}^\dagger \hat{a} ) = e^{-\eta^2/2} \sum_{k=0}^\infty
  \frac{(-1)^k \eta^{2k}}{(k!)^2} \, \hat{a}^{\dagger k} \hat{a}^k ,
\end{equation}
with $\eta$ being the Lamb--Dicke parameter, characterising the
localisation of the atom in the trap. The interaction
Hamiltonian~(\ref{Eq1}) fulfils the condition $[\hat{n}, \hat{H}_{\rm
  int}] \!=\! 0$ of back-action evasion for the vibrational number
operator $\hat{n} \!=\! \hat{a}^\dagger\hat{a}$, so that the latter is
a constant of motion. Consequently, this interaction couples only
between vibronic density matrix elements $\varrho^{mn}_{ij}$ having
the same motional indices $(m,n)$. This renders it possible to
determine the entangled motional number statistics $\varrho^{nn}_{ij}$
of the initial vibronic state by monitoring the dynamics of the
electronic transition $|1\rangle \leftrightarrow |2\rangle$.

One possibility to obtain the matrix elements
${\varrho}_{ij}^{nn}(\alpha)$ is the measurement of the time
dependence of the occupation of state $|2\rangle$. This dynamics is
completely determined by those matrix elements. However, their
reconstruction requires some effort of data analysing by Fourier or
other appropriate techniques, as considered for the Jaynes-Cummings
dynamics \cite{vogel,bardroff-schleich,leibfried,haroche-jcm}.

In the following we will deal with an alternative method that directly
yields the quantities desired in Eq.~(\ref{eq:wigseries}). It is
related to the scheme proposed in Ref.~\cite{matos-vogel-qnd} for the
quantum nondemolition measurement of the motional energy of a trapped
atom.  After a well controlled interaction time of the laser resonant
to the weak electronic transition, a laser pulse is applied on the
strong transition in order to probe for the appearance of resonance
fluorescence. The occurrence of fluorescence detects the atom in the
electronic ground state $|1\rangle$ and its absence in the excited
electronic state $|2\rangle$.  We will focus our attention to the
no-fluorescence events, since they preclude any disturbance of the
motional state via light scattering by the strong transition.
Alternating sequences of light pulses on the weak and the strong
transition allow to reduce the motional state to a Fock state, which
is predetermined by time control of the pulse sequence on the weak
transition.  Eventually, the desired information on the entangled
number statistics ${\varrho}_{ij}^{nn}(\alpha)$ is directly given by
the probability of realizing such a sequence of interaction-probe
cycles, for each predetermined Fock state $|n\rangle$.

Assume that, subsequent to the coherent displacement of the motional
subsystem, the atom is probed for fluorescence on the strong
transition after an interaction time $\tau$ of the laser with the weak
transition.  Provided the atom is detected in the electronic state
$|2\rangle$ (no fluorescence), the density operator of the system
reduces to the (unnormalised) operator
\begin{equation}
  \label{eq:red}
  \hat{\varrho}^{\rm (red)}(\tau) = | 2 \rangle \langle 2 | \otimes
  \hat{\rho}^{\rm (red)}(\tau),
\end{equation}
where $\hat{\rho}^{\rm (red)}(\tau) \!=\! \langle 2|
\hat{\varrho}(\tau) |2\rangle$ is the corresponding (unnormalised)
density operator of the motional subsystem related to the electronic
state $|2\rangle$. In view of the interaction Hamiltonian (\ref{Eq1}),
the diagonal elements of $\hat{\rho}^{\rm (red)}(\tau)$ in
number-state representation are given by
\begin{eqnarray}
  \label{eq:redmot}
  \rho_{nn}^{\rm (red)}(\tau) & = & \varrho^{nn}_{22}(\alpha) \cos^2
  \left( \frac{\Omega_n \tau}{2} \right) + \varrho^{nn}_{11}(\alpha)
  \sin^2 \left( \frac{\Omega_n \tau}{2} \right) \nonumber \\ & & + \,
  {\rm Im} \left[ \varrho^{nn}_{12}(\alpha) e^{-i\varphi} \right] \sin
  \left( \Omega_n \tau \right),
\end{eqnarray}
\narrowtext\noindent where $\Omega_n \!=\! |\Omega| L_n(\eta^2)
\exp(-\eta^2/2)$ are the nonlinear vibronic Rabi frequencies, $L_n(x)$
being Laguerre polynomials.

Let us consider the effect of $k$ of these interaction-probe cycles
with interaction times $\tau_1,\ldots,\tau_k$, each one accompanied by
no fluorescence~\cite{toschek}. The resulting (unnormalised) motional
number statistics is conditioned on the times $\tau_1,\ldots,\tau_k$,
at which the reductions to the state $|2\rangle$ occurred, and reads
as
\begin{equation}
  \label{eq:matcond}
  \rho_{nn}^{\rm (red)}(\tau_k,\ldots,\tau_1) = \prod_{q=2}^k \cos^2
  \left( \frac{\Omega_n \tau_q}{2} \right) \rho_{nn}^{\rm
    (red)}(\tau_1).
\end{equation}
The probability $P(\tau_k, \ldots, \tau_1)$ to obtain such a sequence
of interaction-probe cycles is given by the trace of $\hat{\rho}^{\rm
  (red)}(\tau_k,\ldots,\tau_1)$,
\begin{equation}
  \label{eq:prob}
  P(\tau_k, \ldots, \tau_1) = \sum_{n=0}^\infty \rho_{nn}^{\rm
    (red)}(\tau_k, \ldots, \tau_1),
\end{equation}
and can be experimentally determined by repeating this procedure many
times and counting the number of times one was successful in obtaining
such a sequence.

An adequate choice of the interaction times $\tau_k$ allows to map
$\rho_{nn}^{\rm (red)}(\tau_1)$ onto the probability $P(\tau_k,
\ldots, \tau_1)$.  For mapping the particular element $\rho_{mm}^{\rm
  (red)}(\tau_1)$, it is useful to choose the interaction times
$\tau_2,\ldots,\tau_k$ as
\begin{equation}
  \label{eq:extract}
  \tau_2,\ldots,\tau_k = \frac{2 \pi}{\Omega_m} p, \quad p= 1,2,\ldots
  ,
\end{equation}
where $p$ can be set to a different integer at each interaction cycle.
From Eq.~(\ref{eq:matcond}) it is seen that after an appropriate
number $k\!\ge\!k_{\rm min}$ of interaction-probe cycles
$\rho_{nn}^{\rm (red)}(\tau_k, \ldots, \tau_1)$ reduces to
\begin{equation}
  \label{eq:matcondred}
  \rho_{nn}^{\rm (red)}(\tau_k,\ldots,\tau_1) = \delta_{nm} \,
  \rho_{mm}^{\rm (red)}(\tau_1).
\end{equation}
The other matrix elements ($n \!\neq\! m$) are suppressed by the
product of cosines in Eq.~(\ref{eq:matcond}), provided that different
motional number states are efficiently discriminated by the vibronic
Rabi frequencies $\Omega_n$~\cite{matos-vogel-qnd}.  It is easy to see
[cf.  Eqs~(\ref{eq:prob}), (\ref{eq:matcondred})] that in this case
the probability $P(\tau_k, \ldots, \tau_1)$ reflects the quantity
$\rho_{mm}^{\rm (red)}(\tau_1)$ and that further interaction-probe
cycles will not change anything. Since the interaction times $\tau_k$
are specified from the beginning, the minimum number $k_{\rm min}$ of
cycles needed to complete the mapping process can be evaluated.

The number statistics $\rho_{mm}^{\rm (red)}(\tau_1)$ obtained in this
manner can be used to determine the entangled, motional number
statistics $\varrho_{ij}^{mm}(\alpha)$ of the displaced initial
vibronic state via Eq.~(\ref{eq:redmot}). For this purpose, three
different choices of the first interaction time $\tau_1$ are desired:
\begin{itemize}
\item[(i)] \label{item1} $\tau_1\!=\!0$: After detecting the atom in
  the excited electronic state $|2\rangle$ (no detection of
  fluorescence), $\rho_{mm}^{\rm (red)}(\tau_1)$ is given by
  \begin{equation}
    \rho_{mm}^{\rm (red)}(\tau_1) = \varrho_{22}^{mm}(\alpha).
  \end{equation}
  
\item[(ii)] \label{item2} $\tau_1 \!=\! \pi/\Omega_m$: For this choice
  the diagonal elements of the motional density matrix are given by
  \begin{equation}
    \rho_{mm}^{\rm (red)}(\tau_1) = \varrho_{11}^{mm}(\alpha).
  \end{equation} 
  
\item[(iii)] $\tau_1 \!=\! \pi/(2 \Omega_m)$: In this case one arrives
  at
  \begin{eqnarray}
    \rho_{mm}^{\rm (red)}(\tau_1) & = & \frac{1}{2} \left[
      \varrho^{mm}_{11}(\alpha) + \varrho^{mm}_{22}(\alpha) \right]
    \nonumber \\ & & + {\rm Im} \left[ \varrho^{mm}_{12}(\alpha) e^{-i
        \varphi} \right] .
  \end{eqnarray}
  Choosing two laser phases, $\varphi \!=\! 0$ and $\varphi \!=\!
  -\pi/2$, one can obtain the imaginary and real part of
  $\varrho^{mm}_{12}(\alpha)$, respectively, by subtracting one half
  of the outcomes from (i) and (ii).
\end{itemize}
Eventually, any positive event (series of no fluorescence) recorded in
this manner can be stored in the computer for a direct sampling of the
Wigner-function matrix (to be normalised by the number of trials)
according to Eq.~(\ref{eq:wigseries}).

\section{Numerical simulations}\label{sec:numerics}
A typical example for an entangled vibronic quantum state of a trapped
atom is the state given in Eq.~(\ref{eq:ent}). In Fig.~3 we show a
simulation of all steps of the scheme for the determination of the
Wigner-function matrix of the state $|\psi\rangle$ according to the
method proposed above.
\begin{figure}
  \label{fig3}
  \begin{center}
    \vspace*{-1.5cm}\hspace*{-20mm}\input{sim.pstex_t}\vspace*{-15mm}
  \end{center}
  \caption{Wigner-function matrix $W_{ij}(\alpha)$ of the
    entangled vibronic quantum state $|\psi\rangle$
    [Eq.~(\ref{eq:ent})] with $\beta\!=\!2$. Each matrix element
    $\varrho_{ij}^{mm}(0;\alpha)$ has been numerically sampled with
    $1000$ trials and a sequence of $k\!=\!30$ interaction-probe
    cycles on a $25\!\times\!15$-grid. The Lamb--Dicke parameter is
    $\eta\!=\!0.1$.}
\end{figure}
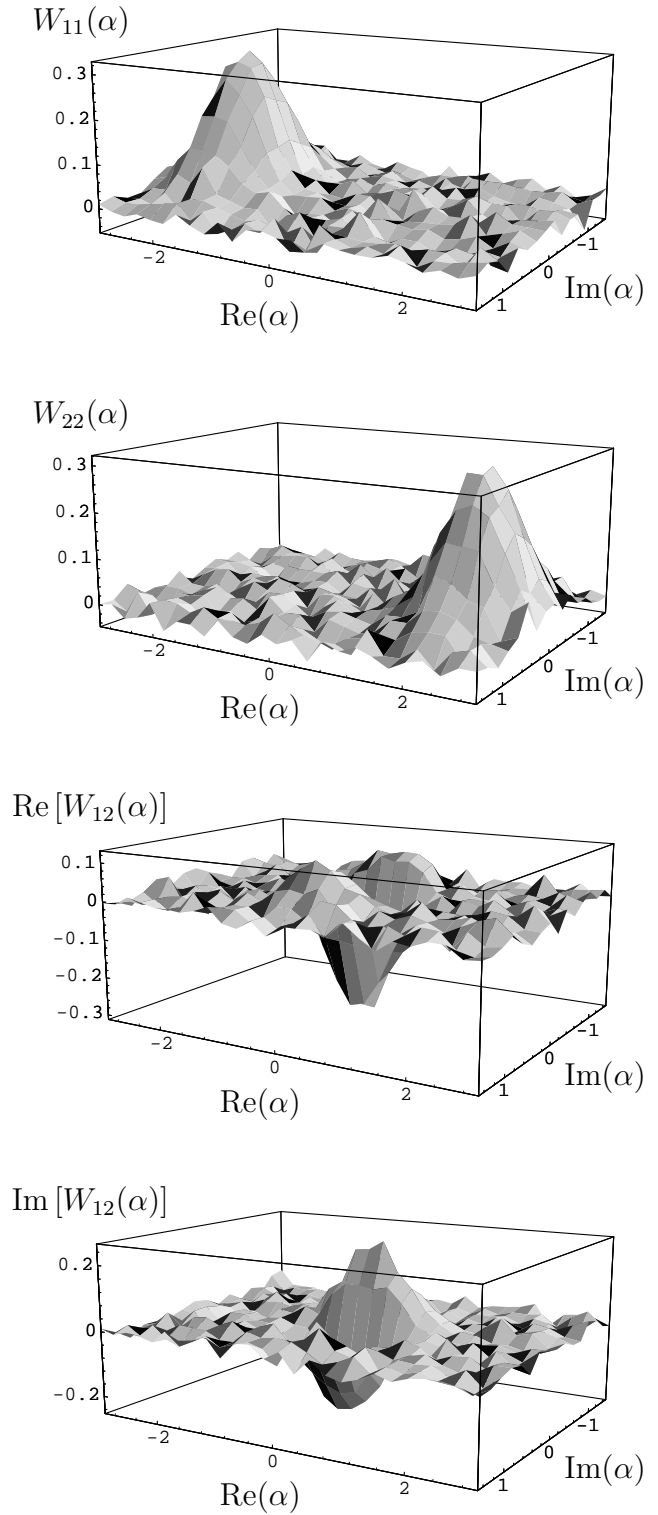
The maximum pulse length of the laser resonant to the weak transition
is of the order of $6\mu$s for a typical Rabi frequency of
$\Omega/2\pi \!=\!  500$ kHz~\cite{monroe}. For interaction times of
this order the electronic relaxation of the weak transition is
negligible. The time needed for the data acquisition of a point in
phase space is approximately $30$ seconds, so that the complete
sampling of one matrix element of the Wigner-function matrix
$W_{ij}(\alpha)$ as shown in Fig.~3 would take about $3$ hours. This
is comparable with the data-acquisition time needed for the
reconstruction of the quantum state of the motional subsystem, by
measurement of the time evolution of the electronic
inversion~\cite{leibfried}. The advantage of our scheme, however,
consists in the fact that the desired number statistics is directly
observed, with no need for inverting systems of linear equations with
noisy data input.

The diagonal elements $W_{22}(\alpha)$ and $W_{11}(\alpha)$ simply
represent the Wigner functions of the coherent states $|\beta\rangle$
and $|-\beta\rangle$ respectively. The off-diagonal elements
$W_{ij}(\alpha)$ contain the more interesting information on quantum
interferences inherent in the entangled state (\ref{eq:ent}).
Comparing with the theoretical result of $W_{ij}(\alpha)$ given in
Fig.~1, a good agreement between theory and simulation is found.  Note
that the entanglement is reflected by the fact that the
Wigner-function matrix is different for different electronic indices.
If one would measure the Wigner function of the motional sub-system,
$W(\alpha) \!=\! \sum_{i} W_{ii}(\alpha)$, the result would be nothing
but an incoherent superposition of two coherent states, completely
concealing the nonclassical character of the atomic state. That is, in
the chosen example the nonclassical nature of the system is manifested
by the entanglement. Our measurement scheme is directly suited to
demonstrate these features.

\section{Summary and conclusion}\label{sec:summary}
In conclusion we have proposed a measurement technique for obtaining
the full information on entangled vibronic states of a trapped atom.
First of all, this requires a characterisation of the full quantum
state of the composed system which can be related to an appropriate
measurement scheme. For this purpose we define a Wigner-function
matrix which has the character of a density matrix with respect to the
electronic degrees of freedom, and of a Wigner function for the
motional subsystem. This Wigner-function matrix is readily related to
the displaced, entangled motional number-statistics of the atom. 

For the determination of the Wigner-function matrix we have presented
a method that allows the measurement of the displaced, entangled
motional number-statistics. It is based on the nonlinearities
appearing in the vibronic coupling of a resonantly driven, weak
electronic transition. This coupling fulfils the back-action evasion
criterion for the number of motional quanta. It eventually allows, in
combination with electronic-state reductions by fluorescence
measurements, the direct statistical sampling of a Wigner-function
matrix displaying the entanglement between the motional and the
electronic degrees of freedom of the atom. The method proposed here
opens new perspectives for experimental demonstrations of fundamental
properties of interacting quantum systems.

\vspace{5mm}

This work was supported by the Deutsche Forschungsgemeinschaft.

\end{document}

%% file: wig.pstex_t
\begin{picture}(0,0)%
\special{psfile=wig.pstex}%
\end{picture}%
\setlength{\unitlength}{0.009375in}%
\begingroup\makeatletter\ifx\SetFigFont\undefined
\def\x#1#2#3#4#5#6#7\relax{\def\x{#1#2#3#4#5#6}}%
\expandafter\x\fmtname xxxxxx\relax \def\y{splain}%
\ifx\x\y   
\gdef\SetFigFont#1#2#3{%
  \ifnum #1<17\tiny\else \ifnum #1<20\small\else
  \ifnum #1<24\normalsize\else \ifnum #1<29\large\else
  \ifnum #1<34\Large\else \ifnum #1<41\LARGE\else
     \huge\fi\fi\fi\fi\fi\fi
  \csname #3\endcsname}%
\else
\gdef\SetFigFont#1#2#3{\begingroup
  \count@#1\relax \ifnum 25<\count@\count@25\fi
  \def\x{\endgroup\@setsize\SetFigFont{#2pt}}%
  \expandafter\x
    \csname \romannumeral\the\count@ pt\expandafter\endcsname
    \csname @\romannumeral\the\count@ pt\endcsname
  \csname #3\endcsname}%
\fi
\fi\endgroup
\begin{picture}(424,980)(5,-160)
\put(394,370){\makebox(0,0)[lb]{\smash{\SetFigFont{12}{14.4}{rm}${\rm Im}(\alpha)$}}}
\put(394,150){\makebox(0,0)[lb]{\smash{\SetFigFont{12}{14.4}{rm}${\rm Im}(\alpha)$}}}
\put(394,-70){\makebox(0,0)[lb]{\smash{\SetFigFont{12}{14.4}{rm}${\rm Im}(\alpha)$}}}
\put(394,590){\makebox(0,0)[lb]{\smash{\SetFigFont{12}{14.4}{rm}${\rm Im}(\alpha)$}}}
\put(224,-85){\makebox(0,0)[b]{\smash{\SetFigFont{12}{14.4}{rm}${\rm Re}(\alpha)$}}}
\put(129, 80){\makebox(0,0)[b]{\smash{\SetFigFont{12}{14.4}{rm}${\rm Im}\left[W_{12}(\alpha)\right]$}}}
\put(224,135){\makebox(0,0)[b]{\smash{\SetFigFont{12}{14.4}{rm}${\rm Re}(\alpha)$}}}
\put(224,355){\makebox(0,0)[b]{\smash{\SetFigFont{12}{14.4}{rm}${\rm Re}(\alpha)$}}}
\put(224,575){\makebox(0,0)[b]{\smash{\SetFigFont{12}{14.4}{rm}${\rm Re}(\alpha)$}}}
\put(124,740){\makebox(0,0)[b]{\smash{\SetFigFont{12}{14.4}{rm}$W_{11}(\alpha)$}}}
\put(124,520){\makebox(0,0)[b]{\smash{\SetFigFont{12}{14.4}{rm}$W_{22}(\alpha)$}}}
\put(129,300){\makebox(0,0)[b]{\smash{\SetFigFont{12}{14.4}{rm}${\rm Re}\left[W_{12}(\alpha)\right]$}}}
\end{picture}

%% file: scheme.pstex_t
\begin{picture}(0,0)%
\special{psfile=scheme.pstex}%
\end{picture}%
\setlength{\unitlength}{0.006875in}%
\begingroup\makeatletter\ifx\SetFigFont\undefined
\def\x#1#2#3#4#5#6#7\relax{\def\x{#1#2#3#4#5#6}}%
\expandafter\x\fmtname xxxxxx\relax \def\y{splain}%
\ifx\x\y   
\gdef\SetFigFont#1#2#3{%
  \ifnum #1<17\tiny\else \ifnum #1<20\small\else
  \ifnum #1<24\normalsize\else \ifnum #1<29\large\else
  \ifnum #1<34\Large\else \ifnum #1<41\LARGE\else
     \huge\fi\fi\fi\fi\fi\fi
  \csname #3\endcsname}%
\else
\gdef\SetFigFont#1#2#3{\begingroup
  \count@#1\relax \ifnum 25<\count@\count@25\fi
  \def\x{\endgroup\@setsize\SetFigFont{#2pt}}%
  \expandafter\x
    \csname \romannumeral\the\count@ pt\expandafter\endcsname
    \csname @\romannumeral\the\count@ pt\endcsname
  \csname #3\endcsname}%
\fi
\fi\endgroup
\begin{picture}(445,381)(80,395)
\put(525,575){\makebox(0,0)[lb]{\smash{\SetFigFont{10}{12.0}{rm}$|2\rangle$}}}
\put(470,395){\makebox(0,0)[lb]{\smash{\SetFigFont{10}{12.0}{rm}$|1\rangle$}}}
\put(290,755){\makebox(0,0)[lb]{\smash{\SetFigFont{10}{12.0}{rm}$|3\rangle$}}}
\end{picture}

%% file: sim.pstex_t
\begin{picture}(0,0)%
\special{psfile=sim.pstex}%
\end{picture}%
\setlength{\unitlength}{0.009375in}%
\begingroup\makeatletter\ifx\SetFigFont\undefined
\def\x#1#2#3#4#5#6#7\relax{\def\x{#1#2#3#4#5#6}}%
\expandafter\x\fmtname xxxxxx\relax \def\y{splain}%
\ifx\x\y   
\gdef\SetFigFont#1#2#3{%
  \ifnum #1<17\tiny\else \ifnum #1<20\small\else
  \ifnum #1<24\normalsize\else \ifnum #1<29\large\else
  \ifnum #1<34\Large\else \ifnum #1<41\LARGE\else
     \huge\fi\fi\fi\fi\fi\fi
  \csname #3\endcsname}%
\else
\gdef\SetFigFont#1#2#3{\begingroup
  \count@#1\relax \ifnum 25<\count@\count@25\fi
  \def\x{\endgroup\@setsize\SetFigFont{#2pt}}%
  \expandafter\x
    \csname \romannumeral\the\count@ pt\expandafter\endcsname
    \csname @\romannumeral\the\count@ pt\endcsname
  \csname #3\endcsname}%
\fi
\fi\endgroup
\begin{picture}(424,980)(5,-160)
\put(394,370){\makebox(0,0)[lb]{\smash{\SetFigFont{12}{14.4}{rm}${\rm Im}(\alpha)$}}}
\put(394,150){\makebox(0,0)[lb]{\smash{\SetFigFont{12}{14.4}{rm}${\rm Im}(\alpha)$}}}
\put(394,-70){\makebox(0,0)[lb]{\smash{\SetFigFont{12}{14.4}{rm}${\rm Im}(\alpha)$}}}
\put(394,590){\makebox(0,0)[lb]{\smash{\SetFigFont{12}{14.4}{rm}${\rm Im}(\alpha)$}}}
\put(224,-85){\makebox(0,0)[b]{\smash{\SetFigFont{12}{14.4}{rm}${\rm Re}(\alpha)$}}}
\put(129, 80){\makebox(0,0)[b]{\smash{\SetFigFont{12}{14.4}{rm}${\rm Im}\left[W_{12}(\alpha)\right]$}}}
\put(224,135){\makebox(0,0)[b]{\smash{\SetFigFont{12}{14.4}{rm}${\rm Re}(\alpha)$}}}
\put(224,355){\makebox(0,0)[b]{\smash{\SetFigFont{12}{14.4}{rm}${\rm Re}(\alpha)$}}}
\put(224,575){\makebox(0,0)[b]{\smash{\SetFigFont{12}{14.4}{rm}${\rm Re}(\alpha)$}}}
\put(124,740){\makebox(0,0)[b]{\smash{\SetFigFont{12}{14.4}{rm}$W_{11}(\alpha)$}}}
\put(124,520){\makebox(0,0)[b]{\smash{\SetFigFont{12}{14.4}{rm}$W_{22}(\alpha)$}}}
\put(129,300){\makebox(0,0)[b]{\smash{\SetFigFont{12}{14.4}{rm}${\rm Re}\left[W_{12}(\alpha)\right]$}}}
\end{picture}

%% file: preprint.bbl
\begin{thebibliography}{99}
\bibitem{footnote1} For a Raman-type excitation the phase $\varphi$ is
  the difference phase of the two Raman beams and can be held very
  stable.
\end{thebibliography}
